\documentclass[fleqn,usenatbib]{mnras}

\usepackage{newtxtext,newtxmath}

\usepackage[T1]{fontenc}

\DeclareRobustCommand{\VAN}[3]{#2}
\let\VANthebibliography\thebibliography
\def\thebibliography{\DeclareRobustCommand{\VAN}[3]{##3}\VANthebibliography}

\usepackage{graphicx}	
\usepackage{amsmath}	

\newcommand{\Msun}{\mbox{M$_{\odot}$}}
\newcommand{\Rsun}{\mbox{R$_{\odot}$}}

\newcommand{\kms}{\mbox{$\mathrm{km\,s^{-1}}$}}
\newcommand{\ms}{\mbox{$\mathrm{m\,s^{-1}}$}}

\newcommand{\Ion}[2]{#1{\,\textsc{#2}}}
\newcommand{\tyc}{TYC\,7218}

\usepackage[usenames,dvipsnames]{xcolor}
\usepackage{color,colortbl}
\usepackage{soul}
\usepackage{graphicx}	
\usepackage{amsmath}	

\usepackage[caption=off]{subfig}
\usepackage[export]{adjustbox}

\title[Bi-modal period distribution of WD+AFGK binaries?]{The White Dwarf Binary Pathways Survey VII: 
Evidence for a bi-modal distribution of post mass transfer systems?
}

\author[F. Lagos et al.]
{
F. Lagos,$^{1,2,3}$\thanks{E-mail: felipe.lagos@postgrado.uv.cl}
M.R. Schreiber,$^{2,4}$
S.G. Parsons,$^{5}$
O. Toloza,$^{2,4}$
B.T. G\"ansicke,$^{6}$
\newauthor
M.S. Hernandez,$^{4}$
L. Schmidtobreick,$^{3}$
D. Belloni$^{4}$
\\
$^{1}$ Instituto de F{\'i}sica y Astronom{\'i}a de la Universidad de Valpara{\'i}so, Av. Gran Breta\~na 1111, Valpara{\'i}so, Chile.\\
$^{2}$ Millennium Nucleus for Planet Formation, NPF, Valpara{\'i}so, Chile\\
$^{3}$ European Southern Observatory (ESO), Alonso de Cordova 3107, Vitacura, Santiago, Chile\\
$^{4}$ Departamento de F\'isica, Universidad T\'ecnica Federico Santa Mar\'ia, Av. España 1680, Valpara\'iso, Chile \\
$^{5}$ Department of Physics and Astronomy, University of Sheffield, Sheffield S3 7RH, UK\\
$^{6}$ Department of Physics, University of Warwick, Coventry, CV4 7AL, UK\\
}

\date{Accepted XXX. Received YYY; in original form ZZZ}

\pubyear{2015}

\begin{document}
\label{firstpage}
\pagerange{\pageref{firstpage}--\pageref{lastpage}}
\maketitle

\begin{abstract}
Binary systems consisting of a white dwarf and a main-sequence companion with orbital periods up to $\approx 100$\,d are often thought to be formed through common envelope evolution which is still poorly understood. To provide new observational constraints on the physical processes involved in the formation of these objects, we are conducting a large-scale survey of close binaries consisting of a white dwarf and an A to K-type companion. 
Here we present three systems with eccentric orbits and orbital periods between $\approx10-42$\,d discovered by our survey. Based on {\textit{HST}} spectroscopy and high angular resolution images obtained with SPHERE-IRDIS, we find that two of these systems are most likely triple systems while the remaining one could be either a binary or a hierarchical triple but none of them is a post common envelope binary (PCEB). The discovery of these systems shows that our survey is capable to detect systems with orbital periods of the order of weeks, but all six PCEBs we have previously discovered have periods below 2.5 d. 
We suggest that the fact that all of the systems we identify with periods of the order of weeks are not PCEBs indicates a transition between two different mechanisms responsible for the formation of very close ($\lesssim 10$\,d) and somewhat wider WD+AFGK binaries: common envelope evolution and non-conservative stable mass transfer.

\end{abstract}

\begin{keywords}
binaries (including multiple): close -- stars: white dwarfs -- techniques: high angular resolution
\end{keywords}



\section{Introduction}

Observational and theoretical population studies of close binaries with white dwarf components are fundamental to progress with our understanding on their formation and evolution, with potentially deep implications for the predicted occurrence rates of systems as important as the progenitors of type Ia supernovae.

Close white dwarfs binaries are believed to form mainly through common envelope (CE) evolution \citep[e.g.][]{webbink84-1,zorotovicetal10-1,ivanovaetal-13} which occurs when the more massive ($\gtrsim1\Msun$) star of an initial main-sequence binary becomes a giant that is filling its Roche lobe, which triggers dynamically unstable mass transfer. 
As the mass transfer time scale is shorter than the thermal time scale of the companion, a common envelope forms engulfing the core of the giant star and the companion.   
Drag forces between the companion and the envelope drive transfer of orbital energy and angular momentum from the orbit of the core of the giant and its companion to the envelope until the latter is fully ejected. The close binary emerging from CE evolution, i.e. the post common envelope binary (PCEB), consists of a white dwarf and a main-sequence companion star also known as the secondary star \citep[e.g.][]{parsons15}.

Despite its importance for close compact binary formation theories, CE evolution is still not fully understood and its physical modelling is frequently simplified by using energy and angular momentum conservation equations involving free parameters that must be fitted through observations. Previous surveys have shown that close white dwarf binaries with M dwarf companions form through CE evolution and led to important constraints on the CE efficiency $\alpha_\mathrm{CE}$, which corresponds to the fraction of orbital energy that is used to unbind the envelope of the white dwarf progenitor. 
Evidence is accumulating that for PCEBs with M dwarf companions $\alpha_\mathrm{CE}$ is rather small, i.e. between 0.2 and 0.3 \citep[e.g.][]{Nelemans2000,zorotovicetal10-1,Nebot2011,Toonen2013}. 
To progress with our understanding of CE evolution it is important to investigate the evolutionary history not only of PCEBs with M dwarfs but over a wide range of companion masses as this may either allow to derive constraints on a universal value for the CE efficiency or indicate the existence of processes that are not included in the simple energy equation.

To that end, we began a large scale survey of close white dwarfs with main sequence star companions of spectral type A, F, G, or K (WD+AFGK binaries). This survey is using the following strategy. As the white dwarf is outshone by the AFGK stars at optical wavelengths, we combined spectroscopic surveys with the Galaxy Evolution Explorer (\textit{GALEX}) database \citep{Bianchi-2011,Bianchi-2017} to identify AFGK stars with ultraviolet (UV) excess as indicative of a potential white dwarf companion \citep{parsons16,rebassa-mansergas17, Ren2020}. We then 
use radial velocity measurements to 
identify close binary systems among our candidates. If radial velocity variations indicate the close binary nature of a given object, we then finally measure the orbital periods of those WD+AFGK candidate systems. We rely on {\textit HST} spectroscopy to confirm that the UV excess is indeed coming from a white dwarf and to measure the mass of the white dwarf.

Following this strategy, we have so far identified the first pre-supersoft X-ray binary system \citep{parsons15}, confirmed that our target selection is reliable and in virtually all cases the UV excess is indeed indicative of a white dwarf \citep{parsons16}, published the first results of our radial velocity campaign \citep{rebassa-mansergas17}, estimated the limited contamination from triple systems motivated by the discovery of the first hierarchical triple system in the survey \citep{Lagos2020}, and identified and characterised 26 new close WD+AFGK systems \citep{Ren2020} through radial velocity variations. 
Most importantly for this work, we measured the periods of six systems and found all of them to be PCEBs with periods below 2.5 days and circular orbits. We reconstructed their evolutionary history and found all of them to be consistent with a small common envelope efficiency \citep[][]{parsons15,Hernandez2020,Hernandez_etal_2022}, in agreement with what was found for PCEBs with M dwarf companions \citep{zorotovicetal10-1,Nebot2011}.

Here we show that our observing strategy does permit the identification 
of systems with longer orbital periods of the order of several weeks. We present the period determination of three systems, i.e.  2MASS\,J06281844-7621467, TYC\,6996-449-1 and TYC\,8097-337-1 (hereafter 2MASS\,J0628, TYC\,6996 and  TYC\,8097 respectively), and find their orbital periods to be between 10 and 42 days and their orbits to be eccentric.  
Using high resolution spectroscopy at optical wavelengths we constrain the stellar properties of the AFGK star, while with high-contrast imaging in the infrared we look for additional companions that could potentially cause eccentric orbits, for instance, through Von Zeipel-Lidov-Kozai oscillations \citep[e.g.][]{Naoz2016}.
We find that one systems (TYC\,6996) is most likely a hierarchical triple system with the white dwarf being the distant tertiary. 
In the case of 2MASS\,J0628, HST spectroscopy does not detect a white dwarf and the UV excess is most likely caused by stellar activity of a lower mass main sequence star.  
For TYC\,8097, given the current available data, we conclude that it could be a triple similar to TYC\,6996, or a binary system with an M or K-type companion. In case of the latter, the UV excess comes either from stellar activity or an active background galaxy.

The fact that all systems from our survey where we measured a period of the order of weeks are not PCEBs, while all systems with periods of less than 2.5 days turned out to be PCEBs, provides crucial constraints on common envelope evolution with potentially deep implications for our understanding of SN\,Ia progenitor channels.

\section{Observations} 
We performed high resolution spectroscopic  
and high contrast imaging observations 
of three targets that have been identified as WD+AFGK candidates by correlating optical surveys and \textit{GALEX} data. In what follows we describe the details of our observational set-ups and the data reduction.

\subsection{Du Pont echelle}

We used the high resolution echelle spectrograph (1 arcsec slit, $R\simeq40\,000$) on the 2.5-m Du Pont telescope located at Las Campanas Observatory, Chile to obtain spectra of our targets. Each science observation was bracketed by ThAr spectra to correct for instrumental drift. However, we place a lower limit on the velocity precision in a single spectrum of $0.5\,{\kms}$ due to the unstable nature of the spectrograph.
Standard image reductions were performed and the spectra optimally extracted and wavelength calibrated using the Collection of Elemental Routines for Echelle Spectra ({\sc ceres}) package \citep{brahm17}. 

\subsection{FEROS}

High resolution spectra were obtained with the FEROS echelle spectrograph ($R\simeq48\,000$) on the 2.2-m Telescope at La Silla, Chile \citep{kaufer98}. FEROS covers the wavelength range from $\simeq3\,500$\,{\AA} to $\simeq9\,200$\,\AA. Observations were performed in Object-Calibration mode where one fibre is placed on the target while the other feeds light from a ThAr+Ne calibration lamp permitting velocity measurements to extremely high precision ($\simeq10\,{\ms}$) and allowed us to correct for instrumental drift throughout the night. FEROS data were reduced using the {\sc ceres} package.

\subsection{CHIRON}

Spectra were also obtained with the CHIRON echelle spectrometer \citep{tokovinin13} on the 1.5-m SMARTS telescope at Cerro Tololo, Chile. We used $3\times1$ binning resulting in $R\simeq40\,000$. CHIRON observations are automatically reduced by the CHIRON team using standard reduction methods. Like the Du Pont observations, we place a lower limit on the velocity precision in a single spectrum of $0.5\,{\kms}$.

\subsection{UVES}

Additional spectra were obtained with UVES \citep{dekker00}, a high resolution echelle spectrograph mounted on the 8.2-m European Southern Observatory Very Large Telescope at Cerro Paranal, Chile. We used the dichroic 1 setup (390+564) with a 0.7 arcsec slit, resulting in $R\simeq50\,000$. The data were reduced using the UVES data reduction pipeline (version 5.8.2). A $0.5\,{\kms}$ lower limit was placed on the velocity precision in a single spectrum.

\subsection{\textit{HST} spectroscopy}
We spectroscopically observed 2MASS\,0628 with the Hubble Space Telescope (\textit{HST}) in order to confirm that the UV excess is due to a white dwarf companion.
We obtained four spectra with the Space Telescope Imaging Spectrograph (STIS) on 2021 April 23 under the program GO 16224 using the G140L grating ($R\sim1\,000$), covering the wavelength range from $\simeq1\,150$\,{\AA} to $\simeq1\,700$\,\AA.
1D spectra were extracted with CALSTIS-6 -- Version 3.4.2 \citep[][]{Hodge_1998} from the bias-subtracted and flat-corrected {\sc flt} files, which were downloaded from the STScI archive. We forced the search to be performed around pixel 400 in the spatial axis, allowing a search of only 30 pixels. The location of the spectra was found to be around the pixel 404. The spectrum was extracted from a box with a width of 11 pixels, while background regions were extracted from boxes above and below the spectrum with sizes of seven pixels. These regions were scaled to the size of the spectrum region.

\subsection{High contrast imaging} 
The three eccentric systems were observed with the high contrast imager VLT/SPHERE \citep{Beuzit_2019} under the programme 100.D-0399. Acquisition of direct imaging with the N-ALC-YJH-S coronagraph was made in the IRDIFS observing mode, which includes IRDIS dual band imaging \citep{Dohlen_2008, Vigan_2010} plus the integral field spectrograph working in Y-J mode \citep{Claudi_2008}. For IRDIS dual-band filters $H2$ ($\lambda_{H2}=1593$\, nm) and $H3$ ($\lambda_{H3}=1667$\,nm) were used. Furthermore, the pupil tracking mode was implemented in order to perform angular differential imaging \citep[ADI,][]{Marois_2006}.

The IRDIS data were first pre-processed (dark background subtraction,
flat-fielding, bad-pixels correction) with the \textsc{vlt/sphere} python package\footnote{\url{https://github.com/avigan/SPHERE}} version 1.4.3 \citep{SPHERE_pipeline}. The frames were recentred based on star centre exposures using the four satellite spots. After pre-processing, and without any post-processing technique to remove speckle patterns produced by the coronograph, we detected one potential companion around 2MASS\,0628 at 0.31 arcsec from the binary. We used the principal component analysis (PCA) algorithm available in the \textsc{Vortex Image Processing} \citep[\textsc{vip},][]{vip_ref, Amara_2012, Soummer_2012} python package to look for fainter companions, finding one companion around TYC\,6996, which is partially visible on the edge of the IRDIS field of view, and two extended sources in TYC\,8097 (see Figure~\ref{fig:sphere_LC}).

\section{Stellar and binary parameters}

From the observations, we derived basic binary and stellar parameters for all three systems. The orbital periods can be derived from our radial velocity measurements, the spectra also provide tight constraints on the nature of the AFGK star in all three systems, and the SPHERE high-contrast imaging provides information on the potential triple nature of our targets. 
 
\subsection{Orbital characterisation}
Radial velocities were computed from all our echelle spectra using cross-correlation against a binary mask representative of a G2-type star (see \citealt{brahm17} for more information). Barycentric correction was applied using {\sc Astropy} \citep{astropy:2018}.
A full list of radial velocity measurements is given in Tables~\ref{tab:velocities1}, \ref{tab:velocities2} and \ref{tab:velocities3} for 2MASS\,J0628, TYC\,8097, and TYC\,6996 respectively in the Appendix.

\begin{figure*}
  \begin{minipage}[t]{0.5\textwidth}
    \subfloat{\includegraphics[width= 1\linewidth,valign=t]{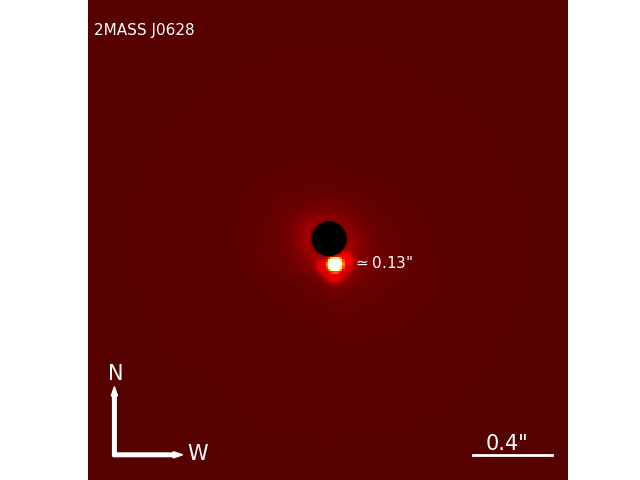}}\\
    \subfloat{\includegraphics[width= 1\linewidth]{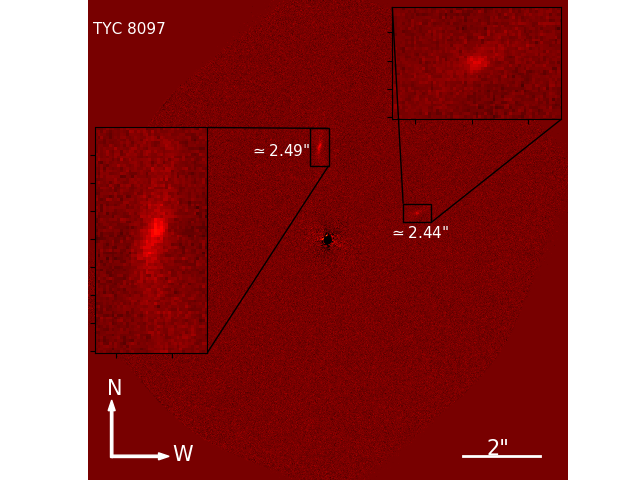}}\\
    \subfloat{\includegraphics[width= 1\linewidth]{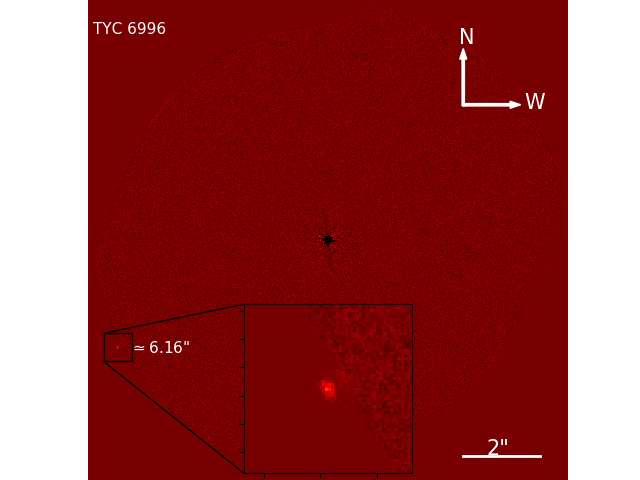}}\\
  \end{minipage}%
  \begin{minipage}[t]{0.5\textwidth}
    \vspace{0.1cm}
    \subfloat{\includegraphics[width= 1.\linewidth,valign=t]{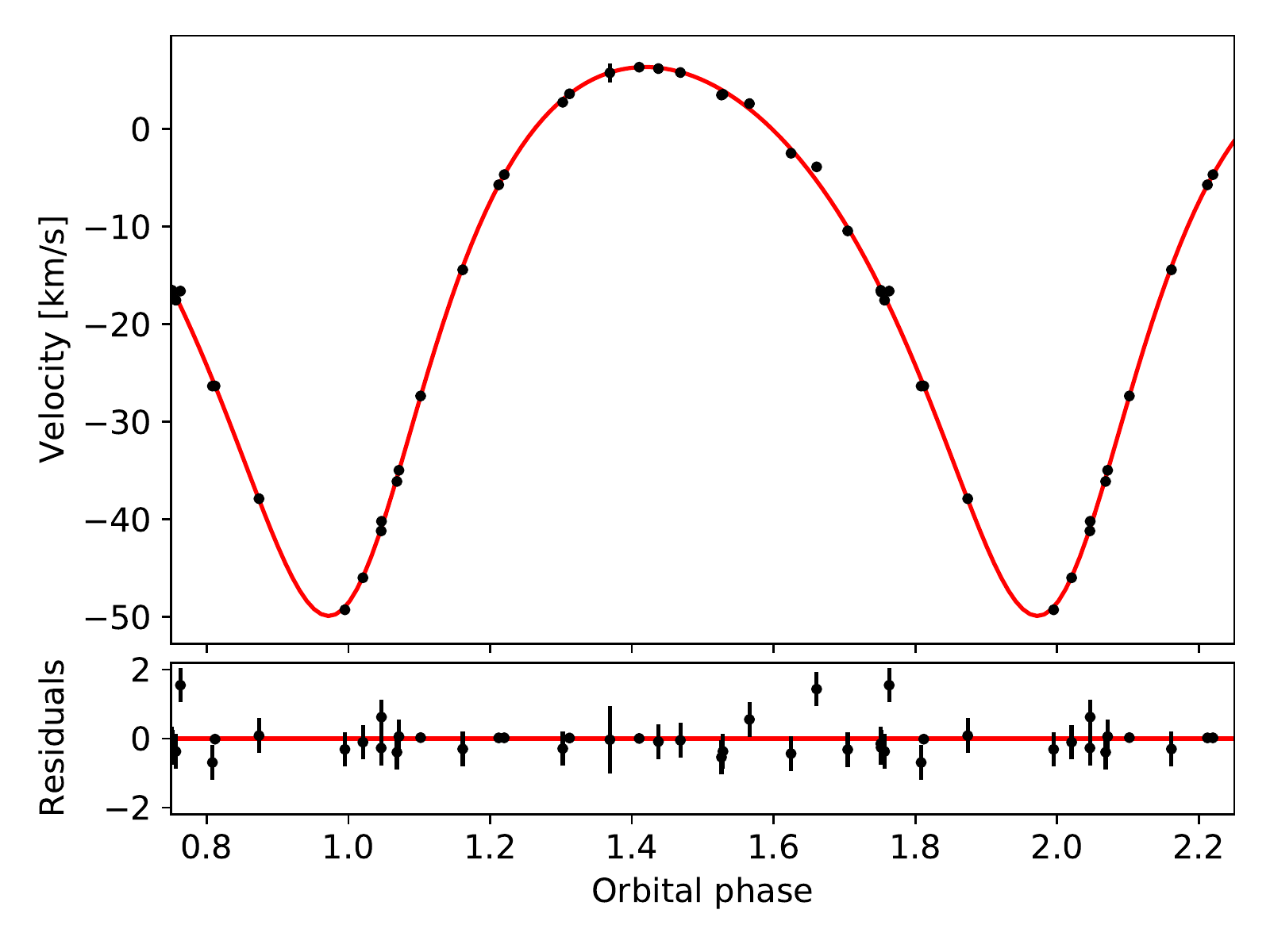}}\\
    
    \vspace{-0.1cm}
    \subfloat{\includegraphics[width= 1.\linewidth]{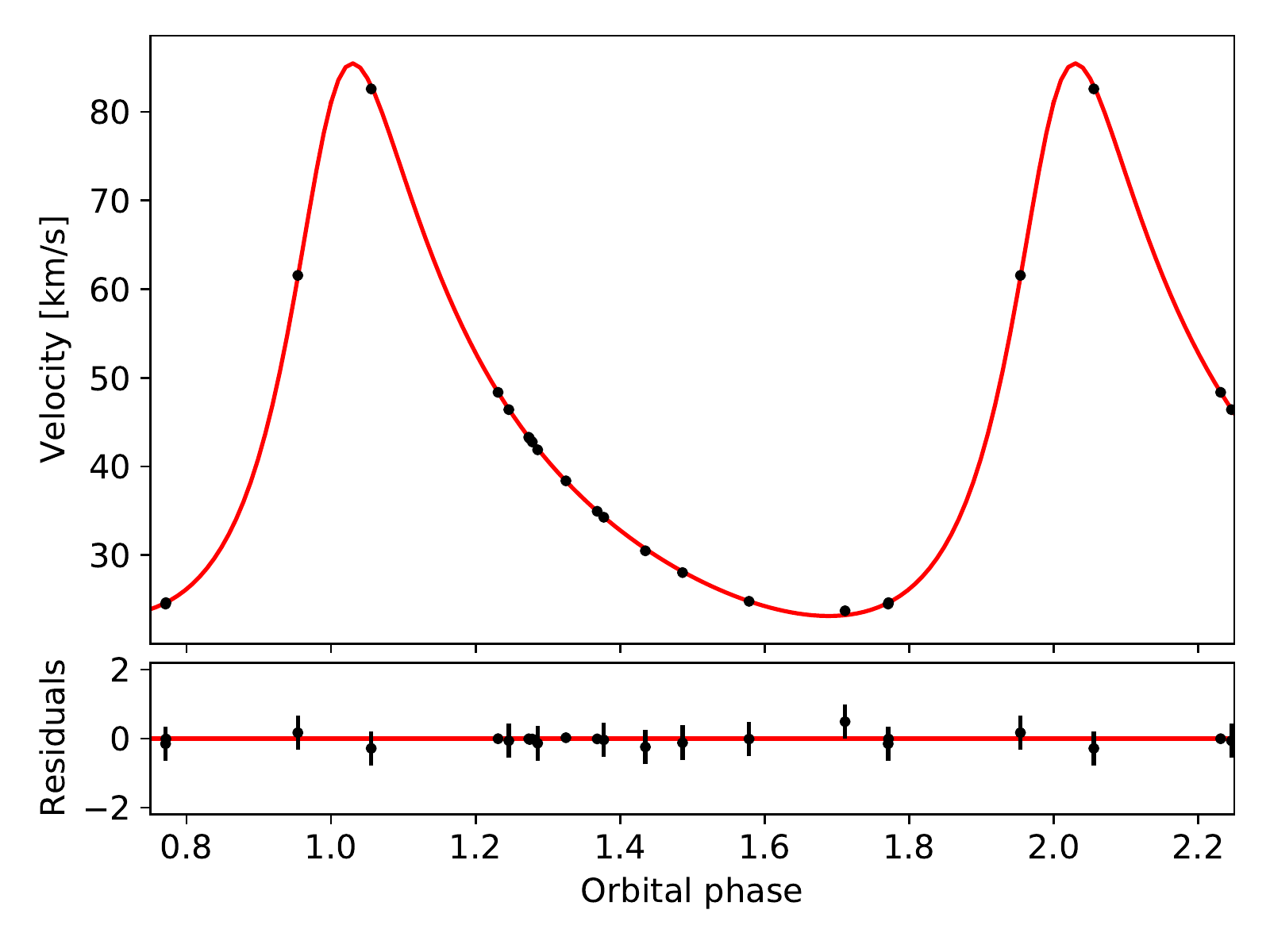}}\\
    
    \vspace{-0.1cm}
    \subfloat{\includegraphics[width= 1.\linewidth]{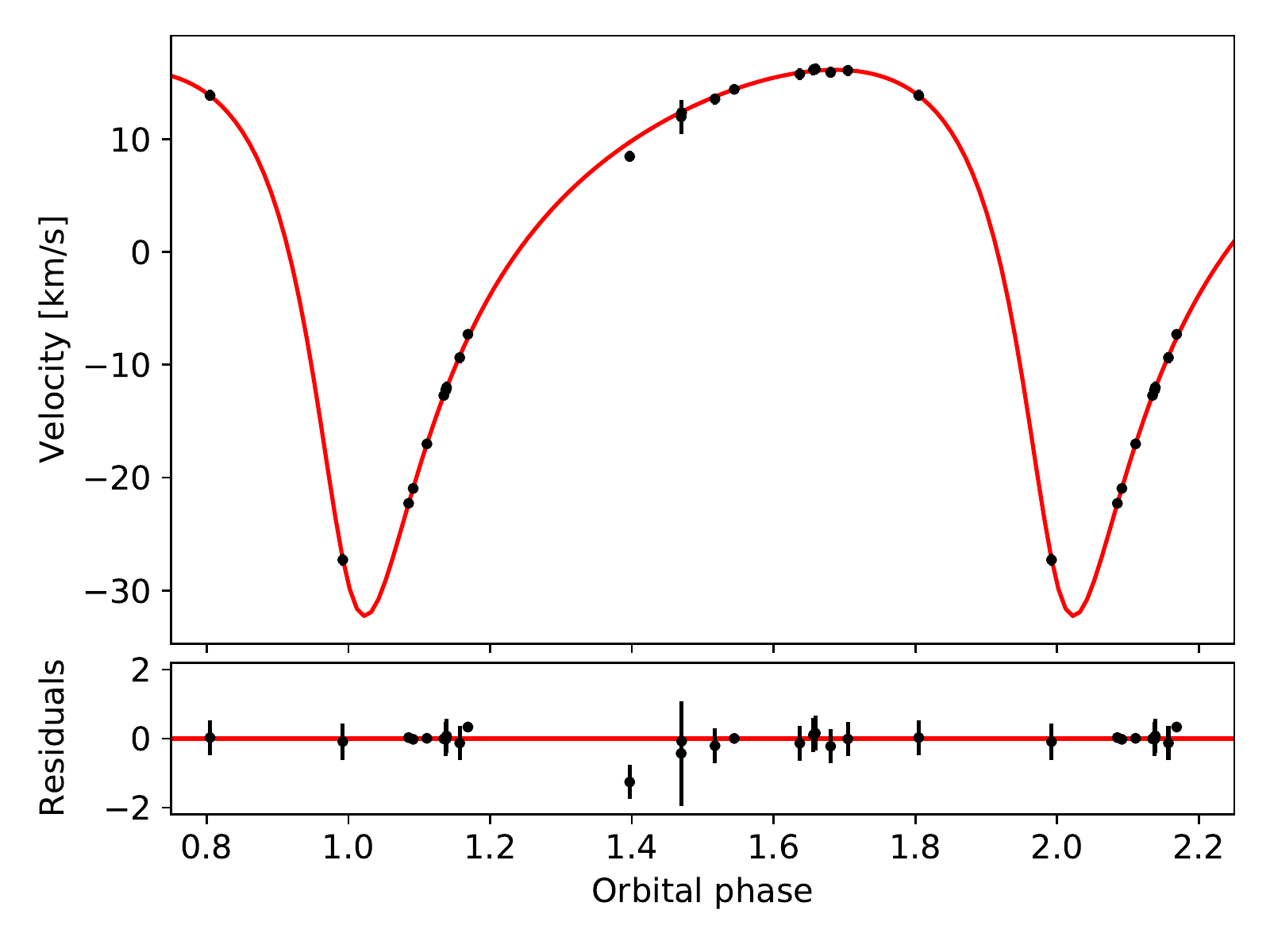}}\\
  \end{minipage}
  \caption{SPHERE/IRDIS $H2$ images (left panel) and radial velocity curves (right panel)  for 2MASS\,0628 (top), TYC\,8097 (middle) and TYC\,6996 (bottom). Black circles in each SPHERE image indicate the position of the coronagraph, while detections are accompanied by its projected separation in arcsec (white numbers). The star dominating at visible light is clearly part of an eccentric binary which excludes these systems to be post common envelope binaries. 2MASS\,J0628 is best explained as a triple systems with at least one active star while TYC\,6996 is also a hierarchical triple with a white dwarf as a tertiary. In the case of TYC\,8097 the origin of the UV excess remains mysterious. The three main candidates to the UV source(s) are the observed background galaxies (in case they are active), stellar activity from the M or K-type companion, or a white dwarf being the third companion in a hierarchical configuration.}
\label{fig:sphere_LC}
\end{figure*}

In order to determine the orbital components of each system ($P_\mathrm{orb}$ - orbital period, $e$ - eccentricity, $K_2$ - radial velocity semi-amplitude, $\gamma$ - systemic velocity, $T_P$ - time of periastron and $\omega$ - argument of periastron) we fitted the radial velocities using {\sc exofast} \citep{eastman13}. For all three targets a clear best fit was found. Uncertainties on the orbital components were determined using the Markov Chain Monte Carlo (MCMC) method \citep{press07} implemented using the python package {\sc emcee} \citep{foreman13}, where the best fit parameters from {\sc exofast} were used as the starting values in the fit. The result of these fits and the radial velocity curves are shown in Table~\ref{table1} and Figure~\ref{fig:sphere_LC} respectively, where from the latter it is clear that the shape of the three radial velocity curves correspond to eccentric orbits.

\subsection{Characterisation of the AFGK stars}

The {\sc ceres} pipeline provides an initial estimate of the stellar parameters of the AFGK star, but only over a narrow parameter range relevant for main-sequence stars. In order to determine more accurate and precise stellar parameters for the AFGK stars in our systems we followed the method outlined in \citet{Hernandez2020}, whereby the high-resolution echelle spectra were fitted using iSpec \citep{blanco14} to determine the effective temperatures {$T_\mathrm{eff,AFGK}$}, surface gravities ($\log g_\mathrm{AFGK}$) and metallicities [Fe/H] ($Z_\mathrm{AFGK}$), which were then used as priors when fitting the optical and infrared spectral energy distribution (SED) of the star in conjunction with the {\it Gaia} EDR3 parallaxes \citep{Gaiaedr3} to determine the radius ($R_\mathrm{AFGK}$). See \citet{Hernandez2020} for a more detailed description of this fitting process. Combining $R_\mathrm{AFGK}$ and $\log{g_\mathrm{AFGK}}$ then yielded the mass of the AFGK star ($M_\mathrm{AFGK}$). The best fit stellar parameters are listed in Table~\ref{table1}.

\subsection{Differential photometry of 2MASS\,0628}
Unlike TYC\,6996 and TYC\,8097, 2MASS\,0628 is the only system which has a point source detection well located in the IRDIS field of view to perform photometry.

We use nine SPHERE/IRDIS flux calibration exposures (from the fits file SPHER.2017-12-30T03:54:37.371) to obtain the differential magnitudes $\Delta H2$, $\Delta H3$ of the companion relative to the instrumental photometry of the central binary. Given that a small fraction of the companion's Airy disk slightly overlaps the first Airy diffraction pattern of the central binary, we use an aperture radius of 4 pixels (i.e., the measured full width at half maximum of IRDIS in $H2$ and $H3$ filters) to avoid as much as possible flux-cross contamination. Sky substraction was performed by calculating the mean value of the pixels located inside an annulus centred at the same angular distance of the companion and width equal to the aperture diameter. The region of the annulus centred at the companion and with length equal to twice the diameter of the aperture was not considered in the calculation.
The final differential magnitudes (i.e. the average of the nine exposures) for filters $H2$ and $H3$ are $\Delta H2=2.17\pm0.03$ and $\Delta H3=2.11\pm0.05$.

\section{The three eccentric systems}

The three targets we present in this paper turned out to have eccentric orbits, which together with \tyc\, \citep{Lagos2020} brings the number of measured eccentric orbits with periods of the order of weeks to four. Given their eccentric nature, these systems are not probably formed though CE evolution \citep[e.g.][]{ivanovaetal-13}. 
One possibility is that, like \tyc, they are triple systems. To elucidate their nature, we have performed detailed follow-up observations to characterise all three systems. 
In what follows we describe each system individually. 

\begin{table}
\caption{Measured mass, radius, effective temperature and metallicity of the AFGK star for each eccentric system. Orbital parameters are obtained from radial velocity measurements of the AFGK star. Photogeometric distances from \citet{Bailer-Jones_2021} are shown with errors based on the 16th and 84th percentiles of the distance posterior distribution. The last two rows represent the most likely source of the UV excess and whether they are triple systems.}
\centering
\tabcolsep=0.05cm
\begin{tabular}{llll} 
\hline
Parameter                                         & TYC 6996         & TYC 8097         & 2MASS J0628      \\ 
\hline
$M_\mathrm{AFGK}$ [$\mathrm{\Msun}$]        &  $1.12 \pm 0.08$ &  $1.43 \pm 0.16$ & $0.99 \pm 0.11$  \\
$R_\mathrm{AFGK}$ [$\mathrm{\Rsun}$]        &  $1.266 \pm 0.020$ &  $1.715 \pm 0.011 $& $0.995 \pm 0.009$  \\
$T_\mathrm{AFGK}$ [K]                       &  $6270 \pm 35$  &  $5730 \pm 40$  & $5540 \pm 50$    \\
$Z_\mathrm{AFGK}$ [[Fe/H]]                          & $-0.31 \pm 0.15$ &  $-0.07 \pm 0.10$ & $-0.07 \pm 0.10$ \\
$P_\mathrm{orb}$ [d]                                 &   $41.9950$    & $20.9850$      & $10.2121$     \\
& $\pm 0.0017$ & $\pm 0.0022$ & $ \pm 0.0013$\\
$e$                                      & $0.497 \pm 0.006$& $0.451 \pm 0.003$& $0.266 \pm 0.002$\\
$K_\mathrm{AFGK}$ [\kms] & $24.21 \pm 0.21$ & $31.05 \pm 0.26$ & $28.11 \pm 0.10$ \\
$\gamma$ [\kms] & $2.54 \pm 0.12$ & $42.17 \pm 0.09$ & $-14.62 \pm 0.02$ \\
$T_\mathrm{P}$ [BJD] & $2\,457\,040.00$ & $2\,457\,423.837$ & $2\,457\,235.406$ \\
& $\pm0.02$ & $\pm 0.001$  & $\pm 0.001$\\
$\omega$ [deg] & $151.9 \pm 0.8$ & $329.0 \pm 0.3$ & $197.7 \pm 0.4$ \\
Distance [pc]            &   $392.1^{+5.3}_{-5.4}$  &     $350.8^{+1.8}_{-1.5}$      & $265.9^{+1.9}_{-1.4}$  \\
triple system            &              most likely               &          inconclusive                &             most likely            \\
UV excess source                              & White dwarf             &     White dwarf         &       Active stars     \\
                              &              &     Active galaxy         &            \\
                              &              &     Active stars         &            \\                         
\hline
\label{table1}
\end{tabular}
\end{table}

\subsection{2MASS\,J06281844-7621467} 
\label{2mass0628}
 
Of all the PCEB candidates observed with \textit{HST}, 2MASS\,J0628 has the faintest \textit{GALEX} $FUV$ magnitude ($m_\mathrm{FUV,GALEX}=21.37 \pm 0.39$) . Based on the orbital properties derived from the radial velocity curve, the minimum mass for the companion to the G-type star is $\approx 0.3\,\Msun$.

\begin{figure}
    \centering
    \includegraphics[width=\columnwidth]{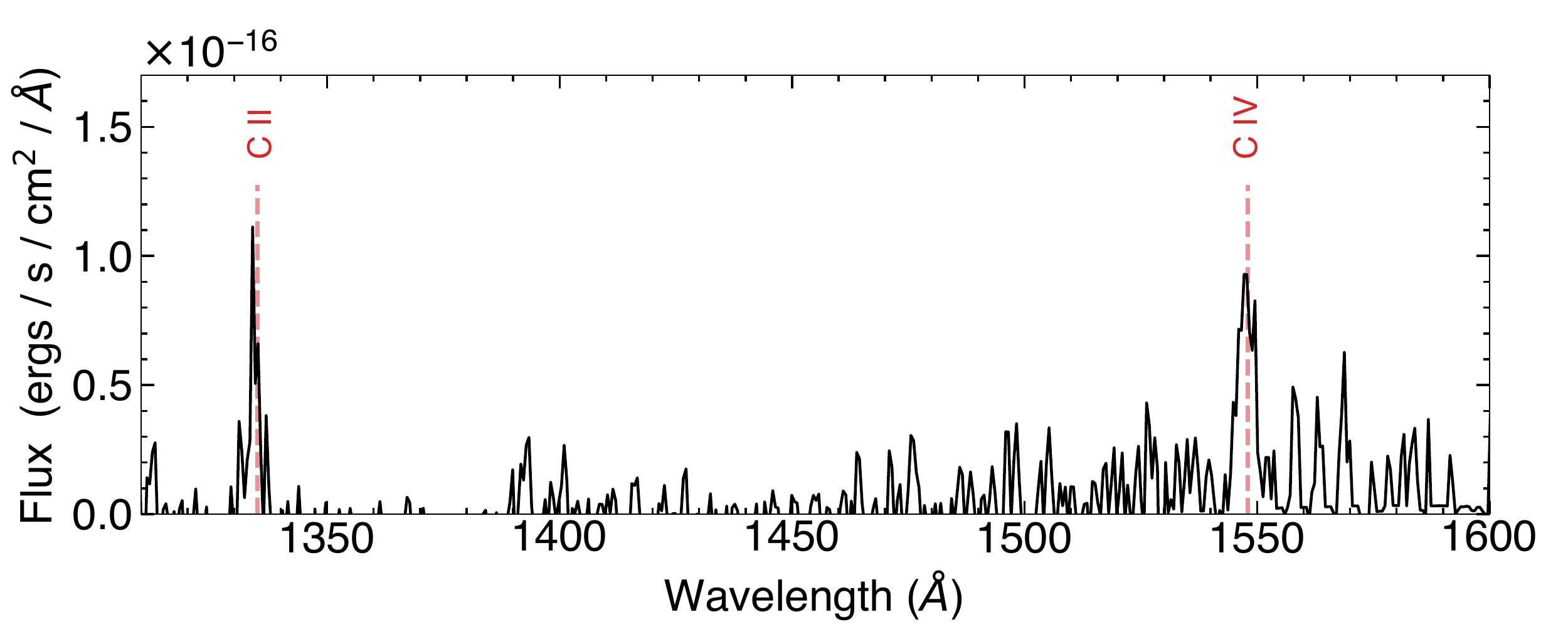}
    \caption{Average \textit{HST}/STIS spectrum of 2MASS\,J0628. A detailed extraction (see text for details) shows clear emission of \Ion{C}{II} and \Ion{C}{iv}. These emission lines may suggest the presence of stellar activity.}
    \label{fig:stis-spectrum}
\end{figure}

The STIS spectra show no evidence for a white dwarf companion or any contribution from a spectral continuum. Averaging the two spectra (from files oe9x08010\_x1d.fits and oe9x08020\_x1d.fits) with the best signal-to-noise ratio, we found two emission features corresponding to \Ion{C}{ii} at 1335\,\AA\ and \Ion{C}{iv} at 1550\,\AA\  (Figure~\ref{fig:stis-spectrum}). Although the integrated flux of this spectrum through the \textit{GALEX} $FUV$ filter is insufficient to explain the observed UV excess, these lines are used as stellar activity tracers and can contribute to the enhanced UV flux produced during transient stages of high activity \citep[e.g.][]{Findeisen2011, Shkolnik2014, Loyd2018}, suggesting that the UV excess may be caused by chromospheric and/or flare activity in at least one star in 2MASS\,J0628.

This idea is further supported by the following reasoning. The $NUV$ flux from the stellar models of G stars do not include chromospheric emission. This explains 
why the models fail to explain the NUV flux of active stars (see the magenta circle in Figure \ref{fig:chi_square} for 2MASS\,J0628). A similar discrepancy between the \textit{GALEX} photometry, the \textit{HST} data, and the stellar model for the AFGK star components has also been observed in one of the six confirmed WD+AFGK PCEBs in our survey: TYC\,110-755-1 \citep[][]{Hernandez_etal_2022}. The white dwarf model that fits best the \textit{HST} spectrum is consistent with the \textit{GALEX} $FUV$ flux, but fails to reproduce the observed \textit{GALEX} $NUV$ flux (see their Figure 9). The authors conclude that chromospheric emission from the G star is the most likely explanation for the flux difference. Based on these two systems, it could be inferred that (1) if the \textit{GALEX} $FUV/NUV$ and \textit{HST}/AFGK model fluxes disagree then variability due to stellar activity might occur in the binary, and (2) if the \textit{GALEX} $FUV$ and \textit{HST} fluxes agree but the \textit{GALEX} $NUV$ flux is larger than the synthetic one, this indicates steady chromospheric emission from the AFGK star.

Furthermore, our SPHERE observations reveal the presence of a companion candidate at an angular separation of 0.13 arcsec, corresponding to a projected separation of $\approx36$\,au using the photogeometric distance from \citet{Bailer-Jones_2021}. To calculate its apparent magnitude in the $H2$ and $H3$ filters we proceed as follows. First, we calculate the synthetic $H2$ and $H3$ magnitudes of 2MASS\,J0628 using the synthetic photometry tool available in the Spanish Virtual Observatory \citep[SVO,][]{SVO,SVO2} and a BT-NextGen \citep{Allard2012} spectral template with effective temperature $5\,500$\,K, surface gravity $\log g=4.5$ and metallicity [Fe/H]=0, i.e., resembling the stellar properties shown in Table \ref{table1}. Then, synthetic magnitudes were reddened using the 3D interstellar dust map from \citet{Lallement19}\footnote{\url{https://astro.acri-st.fr/gaia\_dev/\#extinction}}. The final synthetic magnitudes of the G star in the $H2$ and $H3$ filters are $H2_\mathrm{G}=10.72$, $H3_\mathrm{G}=10.58$, in good agreement with the archival magnitude in the 2MASS $H$ filter of 10.44. With these values, and using the differential photometry, we derived synthetic magnitudes $H2_\mathrm{comp}=12.75$ and $H3_\mathrm{comp}=12.83$ for the companion candidate. Using this information, we estimate the likelihood of the companion to be a background source aligned by chance within an angular distance $\Theta$ following \citet{Brandner2000} as:  
\begin{equation}
  P(\Theta,m_\mathrm{lim})= 1-e^{-\pi \Theta^2 \rho(m_\mathrm{lim})}, 
\end{equation}  
where $\rho(m)$ is the cumulative surface density of background sources down to a limiting magnitude m$_\mathrm{lim}$ (i.e. the magnitude of the detection). In order to calculate $\rho(m_\mathrm{lim})$, we used the Besan\c{c}on galaxy model\footnote{\url{https://model.obs-besancon.fr/}} 
(\citealt{Robin2004}) to generate a synthetic 2MASS $H$ photometric catalogue of point sources within 1 square degree, centred on the coordinates of 2MASS\,J0628. To avoid underestimating the value of P$(\Theta,m_\mathrm{lim})$ we set a limiting magnitude of 16 in the $H$ band, i.e., three magnitudes fainter than the synthetic $H2$ and $H3$ magnitudes of the companion candidate. We found that the probability of the companion to be a background source is 0.001 per cent. It is worth mentioning that this result, based on a synthetic catalogue, is purely statistical and does not confirm the detected object being part of 2MASS\,J0628, but rather allows to strengthen the hypothesis about is triple nature, which needs to be ultimately confirmed via common proper motion through new high contrast observations, as the companion candidate is not resolved by $Gaia$.  

With the evidence presented above, we conclude that 2MASS\,J0628 is most likely a main sequence triple system with at least one active component which produced the UV excess. This system represents the first of the 11 targets in our sample observed with HST that clearly does not contain a white dwarf. Furthermore, by placing 2MASS\,J0628 in the UV colour-temperature diagram of \citet[][their Figure 1]{parsons16}, from which the \textit{PCEB candidate} assignment is given, we might expect that some candidates with  $GALEX$ $FUV-NUV$ colour close or greater than the one of 2MASS\,J0628 ($FUV-NUV\approx\,3.6$) contain UV sources different to white dwarfs, in particular, active stars.

\subsection{TYC 6996-449-1}
\label{tyc_6996}

Previous \textit{HST} observations of TYC\,6996 using STIS were obtained by \citet{parsons16} under program GO 13704. They confirmed the UV excess is due to a white dwarf. However, the flux of the \textit{HST} spectrum for this object falls substantially below its measured \textit{GALEX} magnitudes (see bottom panel of Figure \ref{fig:chi_square}), giving inconsistencies between the distance of the main-sequence star obtained from the RAdial Velocity Experiment \citep[RAVE;][]{kordopatisetal13-1} survey and the white dwarf distance they estimated. As initially suggested by \citet{parsons16}, we show below that this flux difference is because TYC\,6996 is most likely a spatially resolved hierarchical triple with the white dwarf as the tertiary companion. This caused the STIS slit to not be properly centred on the white dwarf losing a substantial part of its UV flux.

\begin{figure}
\begin{center}
\includegraphics[width=\columnwidth]{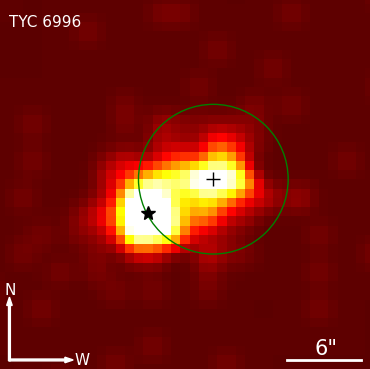}
\caption{\textit{GALEX} FUV image of TYC\,6996 (black cross) and the white dwarf (black star). The SPHERE/IRDIS field of view centred in TYC\,6996 is shown as a green circle. For display purposes, the image has been smoothed with a Gaussian kernel. } 
 \label{fig:galex_6996}
 \end{center}
\end{figure}

Using PCA in the IRDIS science data cube we detected the presence of a potential tertiary companion at $\approx 6.2$ arcsec from TYC\,6996, close to the edge of the IRDIS field of view (see the bottom left panel of Figure \ref{fig:sphere_LC}). This detection is consistent with a UV source detected by \textit{GALEX} at the same coordinates, at $\approx$\,6.1 arcsec from TYC\,6996 as shown in Fig. \ref{fig:galex_6996}, confirming the idea that the white dwarf is either a resolved companion of TYC\,6996 or a background source. According to the \textit{Gaia} EDR3 parallax, the white dwarf is located somewhere between $\approx330-400$\,pc \citep{Bailer-Jones_2021}, in agreement with the estimated distance of TYC\,6996 ($\simeq392$\,pc, see Table \ref{table1}) which supports the triple hypothesis.

In their search of wide companions around stars hosting giant planets and brown dwarfs, \citet{Fontanive2019} found that relative differences of $\leq20$ per cent in the \textit{Gaia} data release 2 parallaxes and in at most one of the proper motion coordinates between the wide companion candidates and the planet-host star may suggest that they are likely part of a hierarchical system. Following this criterion, we calculate the relative differences in proper motion ($\Delta_{\alpha*}$ and $\Delta{\delta}$ ) and parallax ($\Delta\pi$) between the white dwarf and TYC\,6996 using their \textit{Gaia} EDR3 astrometric solutions, finding that $\Delta_{\alpha*}=0.3 \pm 1.1$, $\Delta_{\delta}=22.9 \pm 1$ and $\Delta\pi=8 \pm 14$ per cent. The calculated relative differences meet the criterion that favour the triple nature of TYC\,6996. However, the large error of the white dwarf parallax ($\mathrm{\pi=2.31 \pm 0.35}$) translates into a large uncertainty of $\Delta\pi$, and therefore we cannot discard that its real value could be slightly above the threshold of 20 per cent. However, if the latter was true, and taking into account that the orbital separation of the white dwarf from the inner binary is at least $\approx~2431$~au, the system may also be a triple system that was disrupted due to triple evolution, dynamical instability, galactic tides, encounters with passing field stars or interactions with giant molecular clouds \citep[e.g.][]{Weinberg1987, Correa-Otto2017, Hamers2021, Toonen2021}, in which case the white dwarf could still retain a fraction of the intrinsic proper motion of the former triple.

The presence of triple systems with white dwarfs in the survey has been previously discussed in \citet{Lagos2020}, who found that the white dwarf detected in TYC\,7218 is in fact a tertiary companion to a main sequence binary star. They derived an upper limit of 15 per cent for contaminants (i.e. non-PCEB systems) consisting of hierarchical triples with white dwarfs and M-dwarf companions, of which in 80 per cent of the cases the white dwarf is the third object. We therefore conclude that TYC\,6996 is (or at least was) most likely a triple system, with the close companion to the F-type star being a low mass star with a minimum mass of $\approx0.5~\Msun$ based on its orbital solution.

\subsection{TYC 8097-337-1}
\label{sec:TYC8097}

Based on the orbital solution derived from the radial velocity curve, the minimum mass of the companion to the G-type star is $\approx 0.6 \Msun$.
Since this system was not observed with \textit{HST}, we cannot confirm the presence of a white dwarf, and like 2MASS\,J0628, the companion to the G-type star could, in principle, be an active star producing the observed UV excess. Additionally, IRDIS images reveal the presence of two extended sources at $\approx2.4$\,arcsec from the binary, which are most likely background galaxies that also might be the source of the UV excess. This is the first time in our survey that we have detected not one, but two potential non-stellar UV sources.
 
To quantify the $FUV/NUV$ flux excess produced by the unknown UV source, we perform (as in section \ref{2mass0628}) synthetic $FUV/NUV$ photometry using a BT-NextGen spectral template with effective temperature $T_{\mathrm{eff}}=5\,700$\,K, surface gravity $\log g=4$ and metallicity [Fe/H]=0, according to the stellar properties of the AFGK star given in Table \ref{table1}. While in the $NUV$ band the difference between observed and synthetic magnitudes ($\mathrm{m_{NUV,\textit{GALEX}}=17.06,m_{NUV,synt}=17.22}$) is $\approx$\,0.16\,mag, the difference in the FUV band is $\approx$\,5.6\,mag ($\mathrm{m_{FUV,\textit{GALEX}}=19.93,m_{FUV,synt}=25.52}$). This translates to UV magnitudes of the unknown UV excess source of $\mathrm{m_{NUV}\approx 19.5}$ and $\mathrm{m_{FUV}\approx 19.9}$. 

To investigate if the presence of background galaxies can explain the UV excess, we used the \textit{GALEX} ultraviolet atlas of nearby galaxies of \citet{GildePaz2007}, which mostly contains regular (i.e. non active galactic nuclei, star-forming  or ultraviolet-luminous) galaxies. Since the galaxies in this catalogue have a diameter larger than 1 arcsec, $FUV$ and $NUV$ magnitudes were scaled to such a distance that the projected angular area of the galaxy is roughly equal to that of the two galaxies found with SPHERE, i.e. $\approx$\,$2.3\times10^{-5}\,\mathrm{arcmin^2}$ for a direct comparison with the catalogue. Given that the re-scaled $FUV$ magnitude of the brightest galaxies in the catalogue is $\approx26$\,mag, it seems very unlikely that two regular background galaxies are the unknown UV source. However, if one or both extended sources are active galactic nuclei galaxies, the UV excess can be perfectly explained as these type of galaxies can easily reach $m_\mathrm{FUV}\simeq20$\,mag \citep[e.g.][]{welsh2011}.

In an alternative scenario, the UV excess might be caused by stellar activity.  
We find that using a BT-Nextgen template with T$_{\mathrm{eff}}=4000$ K, $\log g=4.5$ and metallicity [Fe/H]=0 for a companion with 0.6 $\mathrm{\Msun}$, its estimated quiescent FUV magnitude is $\approx$\,39. This result may suggest that it is extremely unlikely that the observed FUV excess is due to stellar activity. However, the spectral template used does not include the chromospheric component, which may contribute $\gtrsim99$ per cent of the total emission in the $FUV$ band and reach flux densities similar to the one observed in TYC\, 8097 \citep[e.g.][]{Stelzer2013}. We therefore conclude that stellar activity cannot be excluded.

\begin{figure}
\begin{center}
\includegraphics[width=\columnwidth]{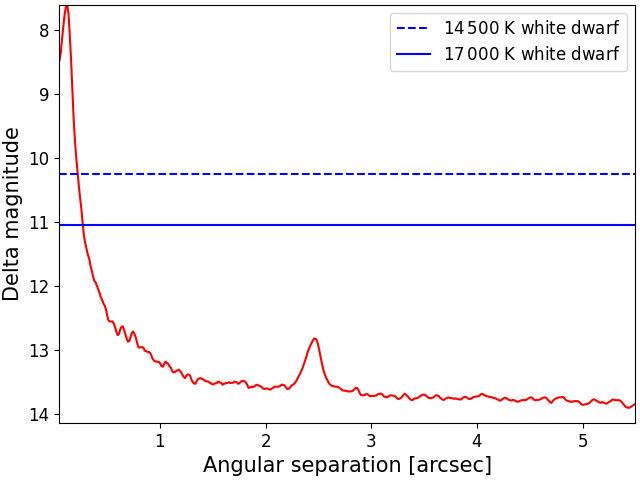}
\caption{5-sigma contrast curve in the H2 filter for a full-frame ADI-PCA using 10 principal components (red line) for TYC\,8097. The solid and dashed blue lines represent the magnitude difference between the AFGK star and the two best-fit models for a white dwarf companion as the source of the UV excess (see the text for more details). The intersection of each blue lines with the red line at $\approx$\,0.27 arcsec represent the limiting angular separation at which the white dwarf would be visible with SPHERE/IRDIS. The peak observed at $\approx 2.4$ arcsec is due to the presence of the background galaxies. Curve generated with \textsc{vip}.} 
 \label{fig:contrast}
 \end{center}
\end{figure}

\begin{figure}
\begin{center}
\includegraphics[width=\columnwidth]{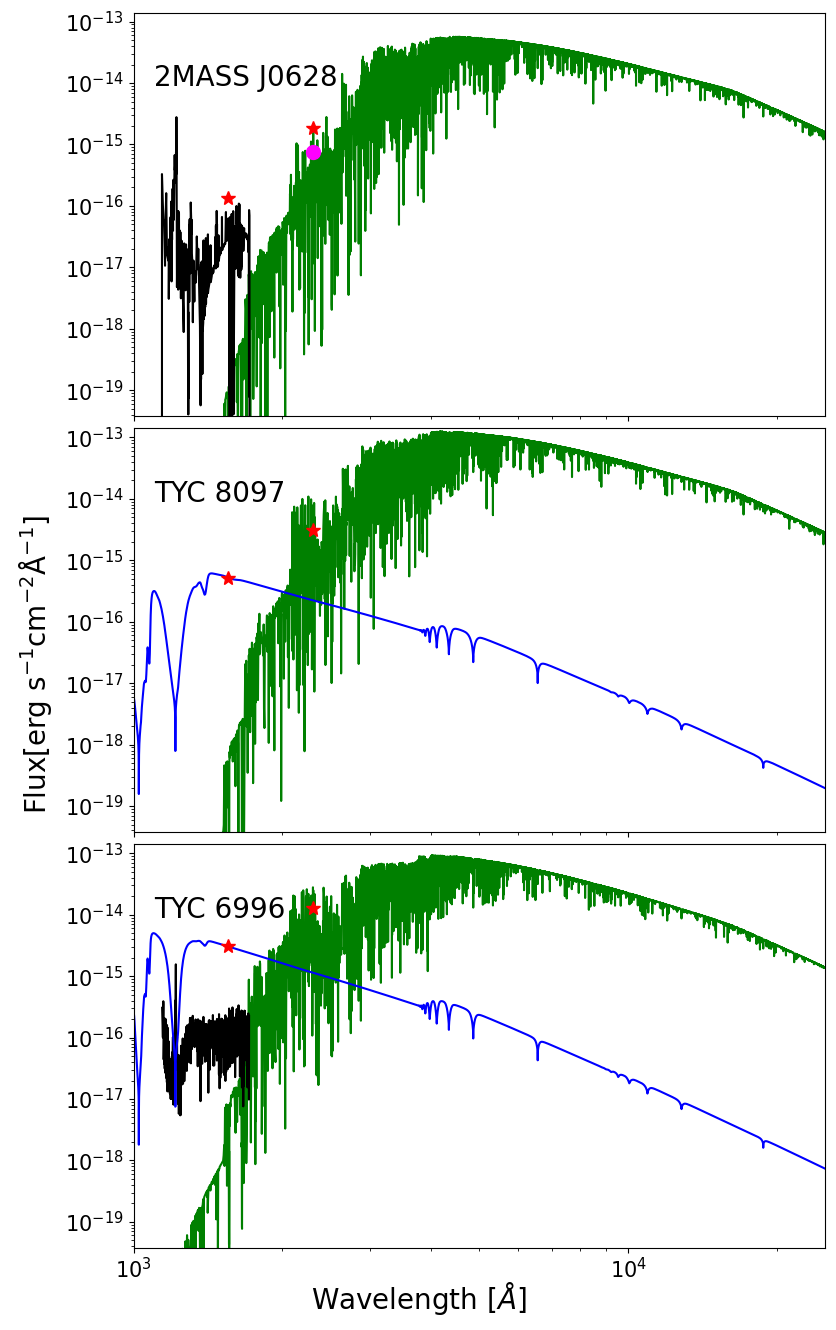}
\caption{Best fit model spectrum of the AFGK star
(green lines) and \textit{GALEX} photometry (red stars) for all the three systems. The HST spectrum of 2MASS\,J0628  (black line) did not confirm the presence of a white dwarf and its flux integrated over the \textit{GALEX} $FUV$ filter is insufficient to explain the observed UV excess. Furthermore, the $NUV$ flux obtained from the model of the G star (magenta circle) is $\approx2.5$ times less than the observed by \textit{GALEX}. We conclude that transient events of stellar activity are responsible for the enhanced \textit{GALEX} UV flux.
For TYC\,8097 we also plot the best white dwarf model fit (T$_{\mathrm{eff}}=14\,500$~K, $\log{g}=8.5$, and M$=0.926\,\mathrm{M}_{\odot}$, blue line) to the observed UV excess. While this is clearly just one possible solution, it illustrates that the UV excess can be explained by the presence of a white dwarf. 
For TYC\,6996 we show the \textit{HST} spectrum (black line) presented also in \citealt{parsons16} from which the presence of the white dwarf was confirmed together with the best white dwarf model fit (T$_{\mathrm{eff}}=16\,500$~K, $\log{g}=8$, and M$=0.616\,\mathrm{M}_{\odot}$) to the \textit{GALEX} photometry. Since the STIS slit was not properly centred on the white dwarf, the observed flux in the \textit{HST} spectrum is smaller than the real one.}
 \label{fig:chi_square}
 \end{center}
\end{figure}

Finally, we test whether a white dwarf may be responsible for the UV excess as follows. Using the white dwarf cooling models\footnote{\url{http://www.astro.umontreal.ca/~bergeron/CoolingModels/}} from \cite{holberg+bergeron06-1,kowalski+saumon06-1, tremblayetal11-2,Bedard2020} we made a chi-square minimisation to the archival $NUV/FUV$ \textit{GALEX} fluxes, leaving as a free parameter the distance of the white dwarf. To include in the minimisation the $FUV/NUV$ flux contribution from the AFGK star we perform synthetic photometry using the same BT-NextGen spectral template described above. Synthetic magnitudes were reddened as explained in Section \ref{2mass0628}. We found that the best fit for the UV excess corresponds to a white dwarf with $\mathrm{T_{eff}=14\,500~K} $, $\mathrm{M=0.926~\Msun}$, $\log g=8.5$ and $\mathrm{R_{WD}=0.97~ R_{\earth}}$, located 356 pc away, in agreement with the distance given in Table \ref{table1}. In addition, we repeat the chi square minimisation, but this time leaving as a free parameter the white dwarf cooling models and keeping fixed the distance at 351 pc, finding that the best fit model corresponds to a white dwarf with $\mathrm{T_{eff}=17\,000} $~K, $\mathrm{M=1.2~\Msun}$, $\log g=9$ and $\mathrm{R_{WD}=0.63~R_{\earth}}$, being consistent with the white dwarf model at 365 pc. 
Figure \ref{fig:contrast} shows that a white dwarf with similar characteristics to our best-fit models would have been detected in our SPHERE/IRDIS images, at least at an angular separation from the AFGK star greater than 0.27 arcsec (projected separation of $\approx95$ au). Given that the eccentricity of TYC\,8097 excludes CE evolution as the formation channel of the binary, the white dwarf might be a tertiary with a close orbit completely hidden by the coronagraph, or we might just have been unlucky and the current location of a wide white dwarf tertiary is just behind the coronagraph. Although this interpretation is also in agreement with the expected fraction of triple systems in the survey with the white dwarf being the third companion, like TYC\,7218 and most likely TYC\,6996, the above scenario must be taken with caution as we only used two photometric points in the chi-square minimisation (Fig. \ref{fig:chi_square}) and only the $FUV$ flux constrains the white dwarf model (making the fit highly degenerate).

We conclude that in the absence of $HST$ observations, the UV source could be either an active galaxy, an active stars, or a white dwarf.

\section{Discussion} 

In the white dwarf binary pathways survey we have so far discovered six short orbital period systems ($P_\mathrm{orb}\lesssim2.5$\,d) with circular orbits. These systems can easily be interpreted as PCEBs. At somewhat longer orbital periods (weeks), we have so far not yet identified a single PCEB but instead four systems, three in this work and one in \citet{Lagos2020}, with eccentric orbits, which are most likely main sequence binaries or the inner binaries of hierarchical triple systems, either with no white dwarf at all or with the white dwarf being the distant tertiary companion candidate. 
In what follows we discuss the implications of this finding for our survey as well as for our understanding of white dwarf binary star formation. 

\subsection{Expected contribution from different types of contaminants}

Despite the effectiveness of the UV excess criterion used to identified white dwarfs with short orbital periods around AFGK-type stars (PCEB candidates), the sample is not free of contaminants, and even if the UV excess comes from a white dwarf, the system may not be formed through CE evolution. In this regard, the orbital characterisation of each candidate, and in particular its eccentricity has become an additional indicator of potential contaminants in the sample. The three eccentric PCEB candidates presented here reveal at least three main type of contaminants: (a) hierarchical triples with the AFGK star being orbited by a M-type star, with a white dwarf as the third object (AFGK/M+WD configuration), (b) binary (triple) systems with the AFGK star being orbited by one (two) M-type star(s) (AFGK/M configuration for binaries and AFGK/M+M for triples), of which one or both are the source of the UV excess due to chromospheric activity, and (c) Back(fore)ground UV sources like white dwarfs or non-stellar sources like galaxies.

The fraction of AFGK/M+WD triples in the survey was derived in \citet{Lagos2020}, ranging from 1 up to 15 per cent of the sample of PCEB candidates. To roughly assess the fraction of type-(b) contaminants, we use an approach similar to that in section 5 of \citet{Lagos2020}. 
For a population composed of $10^7$ single, binary and triple systems, the number of PCEBs with AFGK secondaries is $\approx2\times10^5$ (also including a small fraction formed in hierarchical triples). From the initial mass function of \citet{Kroupa1993}, we estimate that $\approx$3.4 per 
cent of the population are AFGK stars (for the case of single stars) or have an AFGK star as the most massive component (for the case of binary and triple systems). By assuming that 8 per cent of those population of AFGK stars are triples with uniform mass ratio distributions for the inner and outer companions \citep{tokovinin2014}, we found that $\approx$0.06 per cent of the population have triple systems with configuration AFGK/M+M (assuming that M-type stars are those with masses between 0.085 and 0.6 $\Msun$). Compared with the number of PCEBs, the fraction of triples with configuration AFGK/M+M is $\approx3$ per cent. Taking into account the estimated activity lifetimes for M5-M0  and $>$M5-type dwarfs of up to 2 and 8 Gyr respectively \citep{West2011}, and a constant star formation rate, the final fraction of AFGK/M+M systems with at least one active star is reduced to $\approx1$ per cent. Using the same reasoning, but a binary fraction of 33 per cent \citep{tokovinin2014}, we found that the fraction of AFGK/M systems with stellar activity relative to the number of PCEBs is $\simeq9$ per cent. Although the total fraction of type-(b) contaminants ($\approx10$ per cent) is close to the upper limit estimated for AFGK/M+WD contaminants ($\approx15$ per cent), we must recall that the obtained value represents an upper limit, as we assume that all AFGK/M and AFGK/M+M contaminants bypass the UV excess criterion. 

The probability for spurious alignments with back(fore)ground UV sources is very difficult to measure as the fraction of this type of contaminants depends on the type of UV source taken into account and on how crowded the region of the sky around the candidate is. However, since most of the PCEB candidates have galactic latitude and longitude in the range $|b|\gtrsim 30\deg$ and  $200 \lesssim l \lesssim 360 \deg$, i.e. avoiding the crowded regions of the galaxy,  we expect their contribution to be negligible compared with type-(a) and (b) contaminants.

\subsection{Possible transition between systems formed through common envelope evolution and stable non-conservative mass transfer?}

On the one hand,
our survey has successfully identified six PCEBs, all of them with orbital periods less than 2.5 days. On the other hand, the four contaminants identified so far have orbital periods in the range $\mathrm{10\lesssim P_{orb}\lesssim42}$\,d. This shows that our survey is capable to detect wider WD+AFGK binaries thanks to the high precision of our radial velocity measurements. Therefore, if PCEBs with periods longer than 2.5 days would exist in significant numbers, we should have detected them with our observing strategy.

\begin{figure}
\begin{center}
\includegraphics[width=\columnwidth]{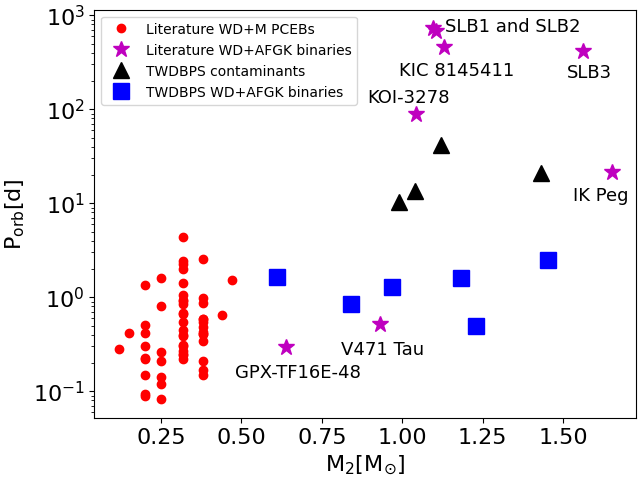}
\caption{Observed relation between orbital period and secondary mass for WD+MS binaries. Red dots are PCEBs with M-type secondaries from \citet{Nebot2011}, magenta stars are WD+AFGK binaries from \citet{Wonnacott1993,O'Brien2001,Kruse_agol2014,Kawahara2018,Masuda2019,Krushinsky2020}, blue squares are PCEBs with AFGK secondaries from The White Dwarf Binary Pathways Survey \citep[TWDBPS,][]{parsons15,Hernandez2020,Hernandez_etal_2022}, black triangles are the contaminants found in our survey presented in this work and in \citet{Lagos2020}.}
 \label{fig:MvsP}
 \end{center}
\end{figure}

In order to find possible explanations for this lack of wider WD+AFGK PCEBs, it is illustrative to look at the secondary mass-period relation of confirmed WD+AFGK binary stars.
In Figure \ref{fig:MvsP} (similar to the Figure 6 of \citealt{Hernandez2020}),
we show WD+MS binaries collected from the literature and those discovered in our survey.
Apart from the six PCEBs identified by our survey, two more WD+AFGK systems have similar short orbital periods, V471\,Tau \citep[$P_{\mathrm{orb}}\approx0.52$\,d;][]{O'Brien2001} 
and GPX-TF16E-48 \citep[$P_{\mathrm{orb}}\approx 0.3$\,d;][]{Krushinsky2020}. 
In the period range where we found four systems that are clearly not PCEBs, only one WD+AFGK binary is known, IK Peg \citep[$P_\mathrm{orb}\approx21.7$\,d, $e\approx0.03$;][]{Wonnacott1993}. At orbital periods longer than $\simeq80$\,d, a population of WD+AFGK binaries has been discovered through the self lensing effect \citep{Kawahara2018,Masuda2019}. We note that a number of white dwarf + A star binaries with similarly long orbital periods are suspected to be present in the Kepler sample statistically analysed by \citet{murphyetal18-1}.

Figure\,\ref{fig:MvsP} indicates a paucity of WD+AFGK binaries in the period range between a few days and more than two months. The only system in this period range is IK\,Peg. Given that four contaminating systems but not a single PCEB in this period range have been discovered by our survey, this paucity is clearly not caused by the fact that longer orbital periods are more difficult to measure by our radial velocity survey. 

Another striking feature illustrated in Figure \ref{fig:MvsP} is the large difference between the results obtained in previous surveys on WD+M binaries \citep{Nebot2011} and the results of our current survey of WD+AFGK. In case of the first, long orbital period (between $80$ and $1000$\,d) systems do not seem to exist, i.e. only PCEBs with periods shorter than a few days and resolved systems with separations exceeding $100$\,au are known \citep{ashleyetal19-1,farihietal10-1}. Our survey of WD+AFGK binaries finds PCEBs with similar characteristics, but there seems to exist a population of systems with orbital periods of several months which is not seen in the M dwarf sample. In what follows we discuss possible implications of these findings.

Binary evolution simulations of PCEBs carried out by \citet{Zorotovic2014} showed that, for systems with $\mathrm{P_{orb}\lesssim10}$ days, the observed relation between the orbital period and the mass of the secondary is well reproduced for a wide range common envelope and recombination energy efficiencies ($\mathrm{0.25\leq \alpha_{CE}\leq 1.0}$ and $\mathrm{0\leq \alpha_{rec}\leq 0.25}$ respectively, see their figure 5). However, WD+AFGK binaries with $\mathrm{P_{orb}\gtrsim100}$ days (such as SBL1, SBL2, SBL3 and KIC 8145411) are only reproduced for models with
significant contributions from recombination energy (i.e. $\mathrm{\alpha_{rec}=0.25}$). Such strong contributions from recombination energy would lead to the prediction of a relatively large number of WD+M PCEBs with periods between 10 days and several months which is clearly not observed. 
If all the systems (except of the contaminants of course) shown in Figure \ref{fig:MvsP} were indeed PCEBs, one would need to explain why nature switches from inefficient envelope extraction for WD+M PCEBs to very efficient envelope extraction including contributions from recombination energy for a fraction of WD+AFGK binaries (i.e from the panels of the left hand side in Fig. 5 of \citet{Zorotovic2014} to the right hand panels of the same figure). Given that the amount of recombination energy that is used to unbind the envelope is neither theoretically nor observationally well constrained, \citep[see\, e.g.][]{ivanova18-1, sokeretal18-1}, we cannot exclude such a dependence of the common envelope efficiencies on the secondary mass. However, we propose an alternative scenario which we think can naturally explain the observations illustrated in Figure \ref{fig:MvsP}. 

We suggest that a different mechanism than CE evolution is responsible for the formation of wide WD+AFKG binaries (with periods of months), and that the period range between 10 and 100 days, populated so far only by our contaminants and two WD+AFGK binaries (IK Peg and KOI-3278), might indeed contain very few post mass transfer white dwarf binaries.
Assuming a universal and low value for the CE efficiency, 
like the one obtained for PCEBs with M-type secondaries ($\mathrm{\alpha_{CE}}\approx0.2-0.3$ \citealt{zorotovicetal10-1}),
PCEBs mostly populate the period domain up to 10 days, predicting only a tail of very few systems with longer periods up to periods of $\mathrm{P_{orb}\approx100}$ days (see figure 2 of \citealt{Zorotovic2014}). CE evolution can explain all the WD+M systems as well as the short orbital period WD+AFGK binaries and probably even IK\,Peg and KOI-3278 as two of the rare longer period PCEBs if a very small fraction of the available recombination energy contributed \citep{Zorotovic2014}. Under these assumptions, however, CE evolution does not predict WD+AFGK with periods of several months as observed and we therefore propose that these systems are not PCEBs.
Instead, dynamically stable non-conservative mass transfer \citep[e.g.][]{Kawahara2018} represents a reasonable candidate to be the mechanism responsible for the formation of these wider WD+AFGK binaries.

The condition for stable mass transfer, under the assumption of adiabatic response of the donor, is generally met when the initial mass ratio $q_i=\mathrm{M_a/M_d}$ between the accretor and the white dwarf progenitor is above a critical mass ratio $q_{\mathrm{crit}}$. For the particular case of conservative mass transfer and donors in the range $\mathrm1{\lesssim \mathrm{M_d} [\Msun] \lesssim6}$, $q_{\mathrm{crit}}\approx1-0.7$ at the termination of the red giant branch \citep{Ge2020}, meaning that stars that are able to evolve off the main sequence within the Hubble time and develop cores (white dwarfs) with masses $\mathrm{\approx 0.5\Msun}$ (as those observed in SBL1, SBL2 and SBL3) can undergo stable mass transfer when the accretor is an AFKG-type star.
If a fraction of the mass lost by the donor escapes from the binary (non-conservative case), then the value of $q_{\mathrm{crit}}$ may be somewhat lower compared to the conservative mass transfer case. 

Indeed, for KIC 8145411, one of the long orbital period post mass transfer WD+AFGK systems, \citet{Masuda2019} suggests a formation channel involving stable (and most likely non-conservative) mass transfer.
Furthermore, non-conservative stable mass transfer, unlike CE evolution, can also explain the long orbital periods of SBL1, SBL2 and SBL3 and KIC\,8145411. As shown in \citet{woods2012}, the final semi-major axis is positively correlated with mass loss rate and initial mass ratio (note that they used $q_i=\mathrm{M_d/M_a}$ instead), being able to increase the orbital period by a factor of five for a binary with $q_i\approx1$, $\mathrm{M_d\approx1.1\Msun}$, and initial orbital period of $\approx100$ days. Finally, if the longer period WD+AFGK stars are descendants from stable mass transfer an obvious explanation for the absence of WD+M binaries in this period range is available: for these systems the mass ratio of the progenitor must have been below 0.5, i.e. the mass of the accretor (the M dwarf) was always too small for stable mass transfer and CE evolution was unavoidable. 

While the outlined scenario is entirely consistent with the currently available observations, as a note of caution, we stress that radial velocity surveys of WD+M binaries have not been sensitive to periods exceeding 10 d \citep{Nebot2011} and that the discovery of self-lensing white dwarf binaries is biased towards systems with larger and brighter companions. A dedicated search of longer period WD+M binaries with a negative result would further confirm our hypothesis.

In addition, given that only ten systems in our survey are well characterised so far (including both PCEBs and contaminants), we admit that despite being plausible, the above outlined interpretation of the observational result remains uncertain. 
By increasing the number of well characterised post mass transfer WD+AFGK binary stars, we can further test our hypothesis and provide crucial constraints on the two formation channels.
This is particularly relevant in the context of SNe Ia progenitors, as WD+AFGK binaries evolving through stable non-conservative mass transfer might experience a subsequent phase of non-stable mass transfer and CE evolution \citep{woods2012}, leaving as end product a double white dwarf binary that may contribute to the SNe Ia rate via the double degenerate scenario.

\section{Conclusion} 

We report the discovery of three WD+AFGK close binary star candidates
whose radial velocity curves revealed unexpected eccentric orbits, 
suggesting that these systems cannot have formed through common envelope evolution. 
By combining high resolution imaging and spectroscopy techniques we confirm the latter, and conclude that two of them are most likely hierarchical triple systems.
2MASS\,J06281844--7621467 has a main-sequence inner binary consisting of G-type primary star and a M or K type companion, which is in turn orbited by another M or K-type star. Since the HST spectrum only shows \Ion{C}{ii} and \Ion{C}{iv} emission lines, the source of the UV excess in this systems is most likely due to one or more active stars. TYC\,6996-449-1 has a main-sequence inner binary, consisting of a F-type primary star and a K or M-type companion, being orbited by an outer white dwarf. The remaining system, TYC 8097-337-1, could be either a binary consisting of a slightly evolved primary G-type star orbited by a K or M-type lower mass companion, or a triple system with a white dwarf as the third companion.

The fact that only contaminants have been found in the orbital period range between 10 and 40 days, may suggests two different mechanisms to form close WD+AFGK binaries: common envelope evolution for orbital periods less than $\approx$\,10 days, and stable non-conservative mass transfer for orbital periods above $\approx$\,100 days. Since binaries evolving through the latter mechanism are expected to end as double white dwarfs after a phase of unstable mass transfer and common envelope evolution, they could play an important role in the formation of supernovae type Ia through the double degenerate scenario. Given the still low number of known WD+AFGK binaries, further characterisation of additional candidates is required to test this hypothesis.

\section*{Acknowledgements}
FL was supported by the ESO studentship and the National Agency for Research and Development (ANID) Doctorado nacional, grant numbers 21211306.
MRS acknowledges financial support from FONDECYT (grant numbers 1181404 and 1221059).
SGP acknowledges the support of the STFC Ernest Rutherford Fellowship. 
OT was supported by a FONDECYT project 321038.
BTG was supported by the UK Science and Technology Facilities Council (STFC) grant ST/T000406/1. 
MSH acknowledges support through a Fellowship for National PhD
students from ANID, grant number 21170070.
This project has received funding from the European Research Council (ERC) under the European Union’s Horizon 2020 research and innovation programme (Grant agreement No. 101020057).
We thank the anonymous referee for his/her very helpful comments and suggestions on the manuscript.
We furthermore thank M\'onica Zorotovic, Robert de Rosa, Anna Pala and Patricia Ar\'evalo
for helpful discussions and suggestions on this work.

\section{DATA AVAILABILITY}
The data underlying this article will be shared on reasonable request
to the corresponding author.



\bibliographystyle{mnras}
\bibliography{example} 




\appendix


\section{Radial velocity measurements}

\begin{table}
 \centering
  \caption{Secondary star radial velocity measurements for 2MASS\,J06281844-7621467}
  \begin{tabular}{@{}lccc@{}}
    \hline
    BJD (mid-exposure) & RV (\kms) & Err (\kms) & Instrument \\
    \hline

2457002.698  & -5.709 & 0.010 & FEROS \\
2457003.719  &  3.619 & 0.010 & FEROS \\
2457004.723  &  6.349 & 0.010 & FEROS \\
2457025.740  &  5.803 & 0.500 & Du Pont echelle \\
2457026.735  &  2.613 & 0.500 & Du Pont echelle \\
2457027.704  & -3.866 & 0.500 & Du Pont echelle \\
2457028.749  &-16.597 & 0.500 & Du Pont echelle \\
2457269.813 &  5.767 & 0.976 & CHIRON \\
2457277.905 &-14.416 & 0.500 & CHIRON \\
2457291.845 &  3.494 & 0.500 & CHIRON \\
2457292.847 & -2.470 & 0.500 & CHIRON \\
2457299.769 &  2.753 & 0.500 & CHIRON \\
2457303.876 &-10.430 & 0.500 & CHIRON \\
2457317.805 &-36.106 & 0.500 & CHIRON \\
2457332.714 &  3.576 & 0.500 & CHIRON \\
2457365.629 &-16.683 & 0.500 & CHIRON \\
2457378.850 &-41.177 & 0.500 & UVES \\
2457386.668  &-26.322 & 0.010 & FEROS \\
2457389.630  &-27.349 & 0.010 & FEROS \\
2457439.599 &-49.261 & 0.500 & UVES \\
2457472.531  & -4.665 & 0.010 & FEROS \\
2457674.735 &-45.979 & 0.500 & UVES \\
2457712.834 &-16.520 & 0.500 & UVES \\
2457723.627 &-26.336 & 0.500 & UVES \\
2457744.722 &-37.877 & 0.500 & UVES \\
2457753.736 &-17.528 & 0.500 & UVES \\
2457760.689 &  6.203 & 0.500 & UVES \\
2457787.586 &-34.963 & 0.500 & UVES \\
2457797.547 &-40.193 & 0.500 & UVES \\

    \hline
    \label{tab:velocities1}
  \end{tabular}
\end{table}

\begin{table}
 \centering
  \caption{Secondary star radial velocity measurements for TYC 8097-337-1}
  \begin{tabular}{@{}lccc@{}}
    \hline
    BJD (mid-exposure) & RV (\kms) & Err (\kms) & Instrument \\
    \hline

2457386.724  & 48.370 & 0.010 & FEROS \\
2457387.640  & 43.159 & 0.010 & FEROS \\
2457387.718  & 42.778 & 0.010 & FEROS \\
2457388.695  & 38.390 & 0.018 & FEROS \\
2457389.603  & 34.955 & 0.010 & FEROS \\
2457471.552  & 43.303 & 0.011 & FEROS \\
2457586.921  & 24.652 & 0.010 & FEROS \\
2457643.893 & 28.042 & 0.500 & UVES \\
2457645.824 & 24.803 & 0.500 & UVES \\
2457655.836 & 82.594 & 0.500 & UVES \\
2457674.689 & 61.560 & 0.500 & UVES \\
2457680.816 & 46.421 & 0.500 & UVES \\
2457684.782 & 30.497 & 0.500 & UVES \\
2457711.565 & 23.734 & 0.500 & UVES \\
2457712.817 & 24.490 & 0.500 & UVES \\
2457723.621 & 41.890 & 0.500 & UVES \\
2457725.542 & 34.282 & 0.500 & UVES \\

    \hline
    \label{tab:velocities2}
  \end{tabular}
\end{table}

\begin{table}
 \centering
  \caption{Secondary star radial velocity measurements for TYC 6996-449-1}
  \begin{tabular}{@{}lccc@{}}
    \hline
    BJD (mid-exposure) & RV (\kms) & Err (\kms) & Instrument \\
    \hline

2456833.851  &-20.966 & 0.022 & FEROS \\
2456835.781  &-12.243 & 0.025 & FEROS \\
2457001.563  &-22.285 & 0.024 & FEROS \\
2457002.645  &-17.019 & 0.026 & FEROS \\
2457003.648  &-12.729 & 0.021 & FEROS \\
2457025.548  & 16.137 & 0.500 & Du Pont echelle \\
2457026.569  & 15.903 & 0.500 & Du Pont echelle \\
2457027.586  & 16.080 & 0.500 & Du Pont echelle \\
2457188.839  & 14.403 & 0.027 & FEROS \\
2457255.781 &-12.018 & 0.500 & CHIRON \\
2457266.632 &  8.456 & 0.500 & CHIRON \\
2457269.684 & 11.963 & 1.515 & CHIRON \\
2457276.712 & 15.748 & 0.500 & CHIRON \\
2457283.761 & 13.860 & 0.500 & CHIRON \\
2457297.742 &-12.244 & 0.500 & CHIRON \\
2457311.706 & 12.334 & 0.500 & CHIRON \\
2457313.682 & 13.547 & 0.500 & CHIRON \\
2457319.630 & 16.206 & 0.500 & CHIRON \\
2457333.623 &-27.295 & 0.527 & CHIRON \\
2457340.552 & -9.372 & 0.500 & UVES \\
2458600.893 & -7.305 & 0.055 & FEROS \\

    \hline
    \label{tab:velocities3}
  \end{tabular}
\end{table}

\bsp	
\label{lastpage}
\end{document}